\documentclass[%
twocolumn,
superscriptaddress,
nofootinbib,
 amsmath,amssymb,
 aps,
]{revtex4-2}
\usepackage{amsmath}
\usepackage{mathptmx}
\usepackage{newtxmath}
\usepackage{hyperref}
\usepackage{graphicx}
\usepackage{dcolumn}
\usepackage{bm}
\usepackage{verbatim}

\newcommand{\xm}{x_\mathrm{min}}
\newcommand{\meanlnx}{\left<\ln x\right>_\Lambda}
\newcommand{\meanx}{\left<x\right>_S}
\newcommand{\hatxm}{\hat x_\mathrm{min}}

\newcommand{\yk}{\left<y_k\right>}
\newcommand{\yksq}{\left<y_k^2\right>}

\begin{document}


\title{Maximum-likelihood fits of piece-wise Pareto distributions with finite and non-zero core}

\author{Benjamin F. Maier}
\affiliation{%
    DTU Compute, Technical University of Denmark, Bygning 321, 2800 Kongens Lyngby, DENMARK
}%

\date{\today}

\begin{abstract}
We discuss multiple classes of piece-wise Pareto-like power law probability density functions $p(x)$ with two regimes, a non-pathological core with non-zero, finite values for support $0\leq x\leq\xm$ and a power-law tail with exponent $-\alpha$ for $x>\xm$. The cores take the respective shapes  (i) $p(x)\propto (x/\xm)^\beta$, (ii) $p(x)\propto\exp(-\beta[x/\xm-1])$, and (iii) $p(x)\propto [2-(x/\xm)^\beta]$, including the special case $\beta=0$ leading to core $p(x)=\mathrm{const}$. We derive explicit maximum-likelihood estimators and/or efficient numerical methods to find the best-fit parameter values for empirical data. Solutions for the special cases $\alpha=\beta$ are presented, as well.
The results are made available as a Python package.
\end{abstract}

\maketitle

\section{Introduction}

Non-negative data that is distributed with a ``heavy tail'' is abundant \cite{newman_networks:_2010,barabasi_network_2016,chotikapanich_modeling_2008,clauset_power-law_2009,broido_scale-free_2019}. There is a whole zoo of theoretical distributions used to describe such data \cite{chotikapanich_modeling_2008,clauset_power-law_2009,broido_scale-free_2019}, one of them the Pareto distribution with a power-law tail $p(x)\propto x^{-\alpha}$ where $\alpha>1$ \cite{chotikapanich_modeling_2008}. This distribution is only non-zero, however, for support $x\geq\xm$ because of the function's pathological behavior at $x=0$. Yet, more often than not, real-world data does not follow a ``lower cutoff'' behavior, instead the empirical probability density function (pdf) reaches finite values for data points that are smaller \cite{clauset_power-law_2009,virkar_power-law_2014}.

Previous work derived and presented maximum-likelihood estimation methods to fit the tail observed in the distributions of empirical data, disregarding values below a threshold $\xm$ \cite{alstott_powerlaw_2014,clauset_power-law_2009}. Doing so is entirely reasonable because the shape of the distribution's tail strongly determines the outcome of dynamical systems \cite{newman_networks:_2010,review-epi-netw} or is an indicator for a system's criticality \cite{schwabl_statistical_2006}. Nonetheless, there might be situations in which the finite, non-zero core of an empirical distribution might be of interest. One such example is the accurate estimation of an empirical contact distribution's first and second moment which are important for epidemic threshold estimations \cite{review-epi-netw} and can both be heavily skewed by outliers. Here, robustly estimating the entire distribution by means of maximum-likelihood estimation may be a more resilient method than simply computing the moments from the data itself.

Therefore, we discuss multiple piece-wise distribution functions that have non-pathological behavior below a threshold, i.e.~have non-zero and finite probabilities for support $0\leq x\leq\xm$ in the following. Each of the distributions takes an additional shape parameter $\beta$ to describe the behavior of the core. For each of the models, we define the log-likelihood and semi-analytical methods to find the parameter values that maximize it given a set of data points. Note that we only present solutions for continuous support $\{x\in\mathbb R: x\geq 0 \}$.

We make the results available as an open-source Python package \cite{github-fincoretails,zenodo-fincoretails}.

\begin{figure*}
    \centering
    \includegraphics[width=\textwidth]{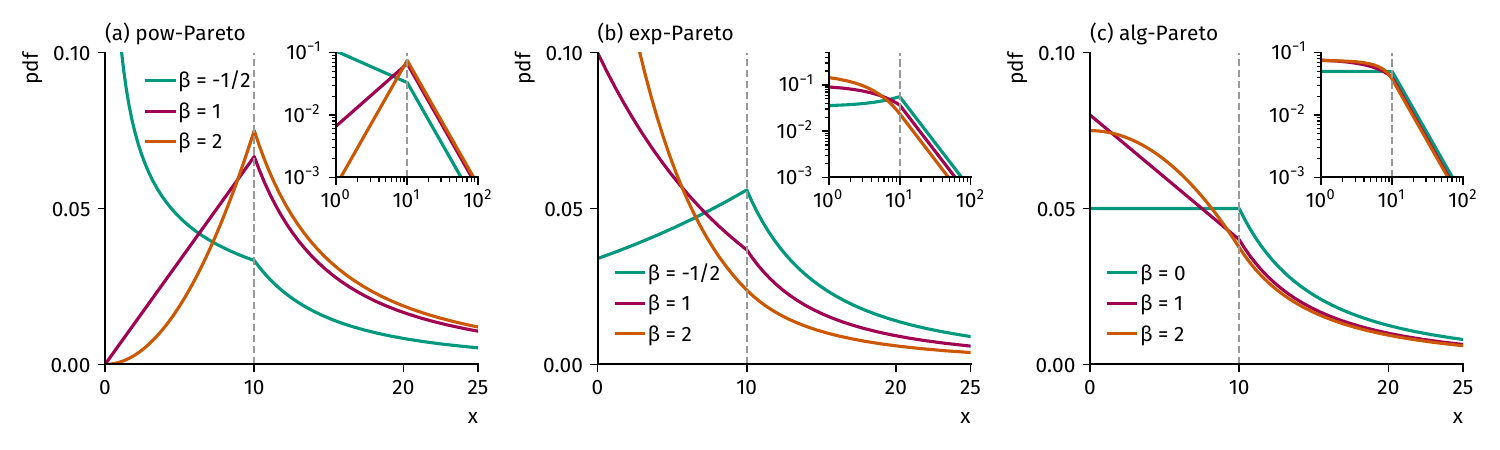}  \caption{Example distributions for the three Pareto-like pdf classes discussed in this paper. Vertically marked is the position of the transition point $\xm$. \textbf{(a)} Power-law-core Pareto distribution Eq.~\eqref{eq:pow-pareto}, discussed in Sec.~\ref{sec:pow-pareto}. \textbf{(b)} Exponential-core Pareto distribution Eq.~\eqref{eq:exp-pareto}, discussed in Sec.~\ref{sec:exp-pareto}. \textbf{(c)} Algebraic-core Pareto distribution Eq.~\eqref{eq:alg-pareto}, discussed in Sec.~\ref{sec:alg-pareto}. Note that $\beta=0$ corresponds to the uniform-core Pareto distribution Eq.~\eqref{eq:uni-pareto}, discussed in Sec.~\ref{sec:unipareto}.} 
    \label{fig:example-plots}
\end{figure*}

\section{Methods}

\subsection{Definitions}
We mainly discuss piece-wise Pareto distributions of the form
\begin{align}
    p(x)=\begin{cases}
            C\gamma(x,\xm,\beta), & 0\leq x\leq\xm\\
            C\left(\xm/x\right)^\alpha, & x>\xm
         \end{cases}
\end{align}
on the support $x\in\{x'\in\mathbb R:x\geq 0\}$ (with a few exceptions where $x>0$). The parameters are bounded as $\alpha>1$ and $\xm>0$. Other bounds will be discussed when appropriate. The respective functions $\gamma$ will be referred to as `core' hereinafter. The normalization constant can be found as 
\begin{align}
    C(\alpha,\xm,\beta) = \left(\frac{\xm}{\alpha-1}+\int\limits_0^{\xm} \gamma(x,\xm,\beta)\right)^{-1}.
\end{align}

Given a set of observations $x_i\in\Omega$ and a threshold $\xm$ we split the data into a set $S\subseteq\Omega$  of observations that are smaller than or equal to $\xm$, and a set $\Lambda\subseteq\Omega$ that contains observations larger than the threshold, i.e. $\Lambda=\{ x_i\in\Omega: x_i > x_m \}$ and $S=\Omega-\Lambda$, with $n_S\equiv|S|$ and $n_\Lambda\equiv|\Lambda|$, as well as $n=|\Omega|=n_S+n_\Lambda$.

The likelihood of a set of observations given a distribution and a parameter set $\{\alpha,\xm,\beta\}$ is therefore
\begin{align}
    \mathcal L(\Omega|\alpha,\xm,\beta) &= C^n(\alpha,\xm,\beta) \prod_{i=1}^{n_\Lambda} \left(\frac{\xm}{x_i}\right)^\alpha
    \prod_{i=1}^{n_S} \gamma(x_i,\xm,\beta).
\end{align}
Consequently, the log-likelihood is given as 
\begin{align}
    \ln\mathcal L &= n\ln C(\alpha,\xm,\beta) -\alpha n_\Lambda
     \left<\ln\left(\frac{x}{\xm}\right)\right>_\Lambda
     \nonumber \\
     &\ \ +n_S
    \big<\ln \gamma(x,\xm,\beta)\big>_S 
    \label{eq:general-log-likelihood}
\end{align}
where we denote as $\left< f(x) \right>_X=(1/n_X)\sum_{i\in X}f(x_i)$ the average over observations in set $X$. Please note that $\left<\ln\left(x/{\xm}\right)\right>_\Lambda = \left<\ln x\right>_\Lambda - \ln\xm$, i.e.~the average can be taken independently from $\xm$ for support regions where the sets $\Lambda$ and $S$ are constant. Furthermore, we have $\left<\ln\left(x/{\xm}\right)\right>_\Lambda>0$ as per the definition of $\Lambda$.

\subsection{General outline of the procedure to find best-fit parameters}

To find parameter values $\hat \alpha$, $\hatxm$, and $\hat \beta$ that maximize the likelihood of a model given a dataset, we proceed as follows.

Typically, we first assume that $\xm$ and $\beta$ are constant and known. Then we solve the equation $\partial\ln\mathcal L /\partial \alpha = 0$, finding $\hat\alpha$. In a next step, we solve $\partial\ln\mathcal L /\partial \beta = 0$ for $\alpha$ to find $\hat\alpha_\beta$. Then, $\hat\beta$ is given as the solution of $\hat\alpha=\hat\alpha_\beta$.

Now, consider the following. Define as $Y=(y_1, y_2, \dots, y_m)$ the ordered tuple of \emph{unique} elements of $\Omega$. For $\xm\in[y_i,y_j)$, the sets $\Lambda$ and $S$ are constant. Therefore, we can use $\partial\ln\mathcal L /\partial \xm = 0$ in concurrence with the previous solutions to find ($\hat\alpha$, $\hatxm$, $\hat\beta$), under the condition that $y_j\leq\hatxm<y_{j+1}$, $\hat\alpha>1$ as well as conditions for $\beta$. In general, it is possible that the maximum likelihood is located at boundary value $\hatxm = y_j$ with $\partial\ln\mathcal L /\partial \xm \neq 0$.

With this in mind, iterate through $\xm=y_1,\dots,y_{m-1}$, construct the respective sets $\Lambda$ and $S$, then compute $\hat\alpha$ and $\hat\beta$ under the assumption that $\xm=y_j=\mathrm{const.}$, afterwards attempt to find a local maximum on the interval $\xm\in[y_j, y_{j+1})$.

\begin{figure*}
    \centering
    \includegraphics[width=\textwidth]{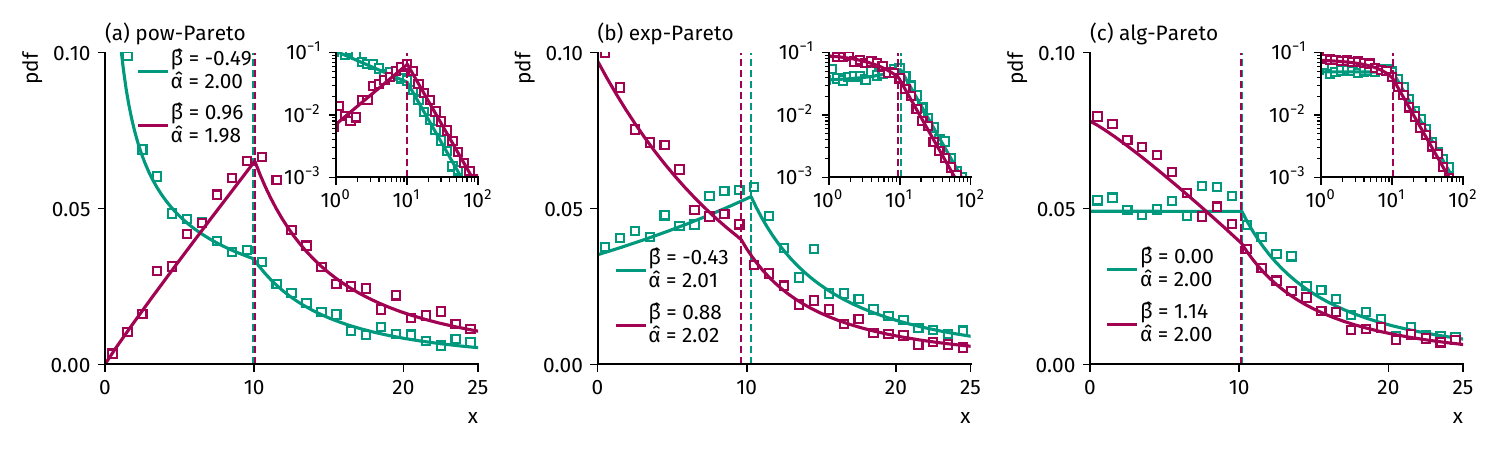}  \\
    \includegraphics[width=\textwidth]{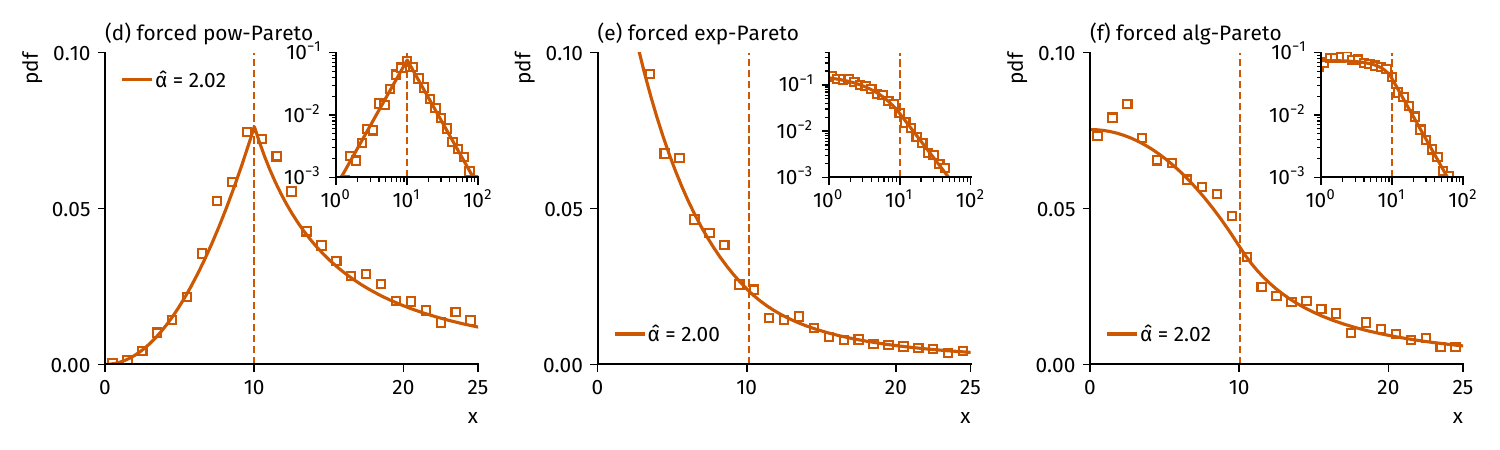}  
    \caption{Random variates sampled from distributions discussed in this paper, their respective empirical pdfs (squares) and maximum-likelihood fits (solid), using $\xm=10$ and $\alpha=2$.
    Vertical lines mark the inferred maximum-likelihood parameter $\hatxm$.
    \textbf{(a)} Power-law-core Pareto distribution Eq.~\eqref{eq:pow-pareto}, discussed in Sec.~\ref{sec:pow-pareto} with $\beta=-1/2$ and $\beta=1$
    \textbf{(b)} Exponential-core Pareto distribution Eq.~\eqref{eq:exp-pareto}, discussed in Sec.~\ref{sec:exp-pareto}
     with $\beta=-1/2$ and $\beta=1$.
     \textbf{(c)} Algebraic-core Pareto distribution Eq.~\eqref{eq:alg-pareto}, discussed in Sec.~\ref{sec:alg-pareto}  with $\beta=0$ and $\beta=1$.
    \textbf{(d)} Forced power-law-core Pareto distribution Eq.~\eqref{eq:force-pow-pareto}, discussed in Sec.~\ref{sec:pow-pareto} 
    with $\beta=\alpha=2$.
    \textbf{(e)} Forced exponential-core Pareto distribution Eq.~\eqref{eq:force-exp-pareto}, discussed in Sec.~\ref{sec:exp-pareto}
    with $\beta=\alpha=2$.
     \textbf{(f)} Forced algebraic-core Pareto distribution Eq.~\eqref{eq:force-alg-pareto}, discussed in Sec.~\ref{sec:alg-pareto}
         with $\beta=\alpha=2$.
     } 
    \label{fig:example-plots}
\end{figure*}
\subsection{Software}

The methods developed herein have been implemented in the Python programming language and are available at \href{https://github.com/benmaier/fincoretails}{https://github.com/benmaier/fincoretails} and \href{https://zenodo.org/record/8349920}{https://zenodo.org/record/8349920}~\cite{github-fincoretails,zenodo-fincoretails}.

\section{Results}
\subsection{Power-law core}
\label{sec:pow-pareto}
\subsubsection{General case}
The ``pow-Pareto'' pdf is given as
\begin{align}
    \label{eq:pow-pareto}
    p(x)=\begin{cases}
            C\left(x /\xm\right)^\beta, & 0\leq x\leq\xm\\
            C\left( \xm /x\right)^\alpha, & x>\xm
         \end{cases}
\end{align}
for $\beta>-1$ and with normalization constant
\begin{align}
    C = \frac{(\alpha -1) (\beta +1)}{y (\alpha +\beta )}.
\end{align}
Note that the pdf has a singularity at $x=0$ for $-1<\beta<0$ but is normalizable nonetheless. We discuss the special case $\beta=0$ in Sec.~\ref{sec:unipareto}.
The log-likelihood is
\begin{align}
\ln\mathcal L &= 
    n \ln \left(\frac{(\alpha -1) (\beta +1)}{y (\alpha +\beta )}\right)
    -\alpha n_\Lambda
     \left<\ln\left(\frac{x}{\xm}\right)\right>_\Lambda
     \nonumber
     \\
    & \qquad
    -\alpha n_S
     \left<\ln\left(\frac{\xm}{x}\right)\right>_S,
\end{align}
with derivative
\begin{align}
\frac{\partial \ln\mathcal L}{\partial \alpha} &= 
\frac{(\beta +1) n}{(\alpha -1) (\alpha +\beta )}-n_\Lambda \left<\ln\left(\frac{x}{\xm}\right)\right>_\Lambda,
    \label{eq:pow-pareto-dLda}
\end{align}
For constant $\beta$ and $\xm$, this yields
\begin{align}
\hat\alpha = \frac{1}{2} \left(1-\beta +(1+\beta) \sqrt{1+\frac{4 n}{(\beta +1)n_\Lambda \left<\ln
   \left(\frac{x}{\xm}\right>_\Lambda\right)}}\right).
   \label{eq:pow-pareto-alphahat1}
\end{align}
The second solution leads to values $\hat\alpha\leq 1$ and can therefore be ignored.

For varying $\beta$ we have
\begin{align}
\frac{\partial \ln\mathcal L}{\partial \beta} &= 
    \frac{(\alpha -1) n}{(\beta +1) (\alpha +\beta )}-  n_S \left< \ln \left(\frac{\xm}{x}\right)\right>_S
    \label{eq:pow-pareto-dLdb}
\end{align}
which yields
\begin{align}
    \hat\alpha &=
    \frac{n+\beta  (\beta +1) n_S \left< \ln \left(\frac{\xm}{x}\right)\right>_S}{n-(\beta +1) n_S \left< \ln \left(\frac{\xm}{x}\right)\right>_S}
   \label{eq:pow-pareto-alphahat2}
\end{align}
and consequently, by equating Eq.~\eqref{eq:pow-pareto-alphahat1}  and Eq.~\eqref{eq:pow-pareto-alphahat2} the two solutions
\begin{align}
\hat\beta_+ &= -1+
    \frac{n}{n_S \left< \ln \left(\frac{\xm}{x}\right)\right>_S+\sqrt{n_\Lambda n_S \left< \ln \left(\frac{\xm}{x}\right)\right>_S
    \left<\ln\left(\frac{x}{\xm}\right)\right>_\Lambda
    }}\\
\hat\beta_- &= -1+
    \frac{n}{n_S \left< \ln \left(\frac{\xm}{x}\right)\right>_S
    -
    \sqrt{n_\Lambda n_S \left< \ln \left(\frac{\xm}{x}\right)\right>_S
    \left<\ln\left(\frac{x}{\xm}\right)\right>_\Lambda
    }}
\end{align}
Only $\hat\beta_+$ generally meets the condition $\beta > -1$, nonetheless, it is computationally cheap to check both solutions.

For regions $\xm\in[y_j,y_{j+1})$, the sets $\Lambda$ and $S$ are constant. We have
\begin{align}
\frac{\partial \ln\mathcal L}{\partial \xm} &= 
-\frac{n-\alpha  n_\Lambda+\beta  n_S}{\xm}
\end{align}
and therefore
\begin{align}
\hat\alpha=
\frac{n(\beta+1)-\beta  n_\Lambda}{n_\Lambda}
\end{align}
where we used $n_S=n-n_\Lambda$. We put this in Eq.~\eqref{eq:pow-pareto-dLda} and solve for $\ln\xm$ to find
\begin{align}
    \ln\hatxm &= \frac{n_\Lambda}{(\beta +1) (n-n_\Lambda)}+\left<\ln x\right>_\Lambda.
\end{align}
Both these results only depend on $\beta$, which is why we can use them in Eq.~\eqref{eq:pow-pareto-dLdb} to find
\begin{align}
    \hat\beta &= -1 +
    \frac{n_\Lambda-n_S}
    {n_S
   (\left<\ln x\right>_S-\left<\ln x\right>_\Lambda)}.
\end{align}
Note that because $\left<\ln x\right>_S<\left<\ln x\right>_\Lambda$, the condition $\beta>-1$ is only met if $n_S>n_\Lambda$.
As described in the Methods section, we can iterate through the ordered tuple $Y$ to find the maximum as either of the boundaries of an interval $[y_j,y_{j+1})$ or within the interval. 

\subsubsection{Special case $\beta=\alpha$}
The ``forced pow-Pareto'' pdf is given as
\begin{align}
    \label{eq:force-pow-pareto}
    p(x)=\begin{cases}
            C\left( x /\xm\right)^\alpha, & 0\leq x\leq\xm\\
            C\left(\xm /x\right)^\alpha, & x>\xm
         \end{cases}
\end{align}
with 
\begin{align}
    C = \frac{\alpha^2 -1}{2y \alpha}.
\end{align}
The log-likelihood is
\begin{align}
\ln\mathcal L &= 
    n \ln \left(\frac{\alpha^2 -1}{2y \alpha}\right)
    -\alpha n_\Lambda
     \left<\ln\left(\frac{x}{\xm}\right)\right>_\Lambda
    -\alpha n_S
     \left<\ln\left(\frac{\xm}{x}\right)\right>_S,
\end{align}
with derivative
\begin{align}
\frac{\partial \ln\mathcal L}{\partial \alpha} &= 
\frac{\left(\alpha ^2+1\right) n}{\alpha ^3-\alpha }
    - n_\Lambda
     \left<\ln\left(\frac{x}{\xm}\right)\right>_\Lambda
    - n_S
     \left<\ln\left(\frac{\xm}{x}\right)\right>_S, 
   \label{eq:force-pow-pareto-dLda}
\end{align}
While this equation is in principle solvable, the three solutions are not very insightful. We can compute the second derivative 
\begin{align}
\frac{\partial^2 \ln\mathcal L}{\partial \alpha^2} &= 
    -\frac{\left(\alpha ^4+4 \alpha ^2-1\right) n}{\alpha ^2 \left(\alpha ^2-1\right)^2}
\end{align}
And use both in Newton's method to find $\hat\alpha$ as the root of Eq.~\eqref{eq:force-pow-pareto-dLda}.

For regions $\xm\in[y_j,y_{j+1})$, the sets $\Lambda$ and $S$ are constant. We have
\begin{align}
\frac{\partial \ln\mathcal L}{\partial \xm} &= 
-\frac{n-\alpha  n_\Lambda+\alpha  n_S}{\xm}
\end{align}
and therefore
\begin{align}
\hat\alpha=
\frac{n}{n_\Lambda-n_S}.
\end{align}
Note that (i) for $n_\Lambda=n_S$ there is no zero, (ii) for $n_S > n_\Lambda$ the solution for $\hat\alpha$ becomes negative and is therefore out of our range. We put this result in Eq.~\eqref{eq:force-pow-pareto-dLda} and solve for $\ln\xm$ to find
\begin{align}
    \ln\hatxm &=
1
-\frac{n^2}{2 n_\Lambda n_S}+\frac{n_S \left<\ln x\right>_S-n_\Lambda \left<\ln x\right>_\Lambda}{n_S-n_\Lambda}.
\end{align}
We can proceed as above, iterating through the ordered tuple $Y$ to find the maximum as either of the boundaries of an interval $[y_j,y_{j+1})$ or within the interval.

\subsection{Exponential core}
\label{sec:exp-pareto}
\subsubsection{General case}
The ``exp-Pareto'' pdf is defined as
\begin{align}
    \label{eq:exp-pareto}
    p(x)=\begin{cases}
            C\exp\Big[-\beta(x/\xm-1)\Big], & 0\leq x\leq\xm\\
            C\left(\xm /x\right)^\alpha, & x>\xm
         \end{cases}
\end{align}
for $\xm>0$, $\beta\neq0$, and $\beta\neq\alpha$ and with normalization constant
\begin{align}
    C = \frac{\beta(\alpha-1)}{\xm \Big((\alpha-1)(e^\beta -1)+\beta\Big)}.
\end{align}

The log-likelihood is given as
\begin{align}
\ln\mathcal L &= n\left[\ln\beta +\ln(\alpha-1) -\ln\xm -\ln\Big((\alpha-1)(e^\beta -1)+\beta\Big)\right]  \nonumber \\
& \ \  - \alpha n_{\Lambda}\left<\ln\frac x \xm\right>_\Lambda - \beta n_{S} \left(\frac{\meanx}{\xm} - 1\right) 
\end{align}
with derivative
\begin{align}
    \frac{\partial\ln\mathcal L}{\partial \alpha} &= 
\frac{n}{\alpha-1} -\frac{n\left(e^\beta -1\right)}{(\alpha-1)(e^\beta -1)+\beta}     - n_{\Lambda}\left<\ln\frac x \xm\right>_\Lambda,
\end{align}
such that for constant $\beta$ and $\xm$ we find
\begin{align}
    \hat\alpha &= 
    1- \frac{\beta}{2 \left(e^\beta -1\right)}\left(
    1-
    \sqrt{1+ \frac{4 n \left(e^{\beta}-1\right)}{\beta n_\Lambda\left<\ln(x/\xm)\right>_\Lambda} }
    \right).
\label{eq:genexp-hatalpha-1}
\end{align}
The second solution to $\alpha$ would give $\hat\alpha<1$ which he have ruled out (note that $\beta/(e^\beta -1)>0$ for negative and positive beta---the second solution has a plus sign preceding the root, which would yield $\hat\alpha<1$). Moreover, we have
\begin{align}
    \frac{\partial\ln\mathcal L}{\partial \beta} &= 
    -\frac{(\alpha -1) \left(e^{\beta } (\beta -1)+1\right) n}{\beta  \left((\alpha -1) e^{\beta }-\alpha +\beta +1\right)}-n_{S} \left(\frac{\meanx}{\xm} - 1\right).
    \label{eq:genexp-dLdb}
\end{align}
For a fixed value of $\xm$, one may find $\hat\beta$ numerically as the solution to $\partial\ln\mathcal L/\partial\beta\big|_{\alpha=\hat\alpha} = 0$.

If we allow $\xm$ to vary, too, we have
\begin{align}
    \frac{\partial\ln\mathcal L}{\partial \xm} &=
    \frac{-n\xm+n_\Lambda \xm \alpha + n_S\meanx\beta}{\xm^2}
    \label{eq:genexp-dLdxm}
\end{align}
for values of $\xm$ where the sets $\Lambda$ and $S$ are constant. Setting Eqs.~\eqref{eq:genexp-dLdb} and ~\eqref{eq:genexp-dLdxm} equal to zero and solving for $\xm$ gives the conditions
\begin{align}
    \hatxm &= \frac{\beta  n_S \meanx \left((\alpha -1) e^{\beta }-\alpha +\beta +1\right)}{\beta  n_S \left((\alpha -1) e^{\beta }-\alpha +\beta +1\right)-(\alpha -1)
   \left(e^{\beta } (\beta -1)+1\right) n},\\
   \hatxm &= \frac{\beta  n_S \meanx}{n-\alpha n_\Lambda}.
   \label{eq:genexp-hatxm-2}
\end{align}
Equating both and using the identity $n=n_S+n_\Lambda$, we can find $\hat\alpha$ as 
\begin{align}
\hat\alpha = 1 + \frac{n_S}{n_\Lambda}\frac{\beta}{e^\beta -1}.
\label{eq:genexp-hatalpha-2}
\end{align}
The second solution gives $\alpha=\beta$, which we have ruled out. The solution above (Eq.~\eqref{eq:genexp-hatxm-2}) gives
\begin{align}
    \hatxm =\meanx\frac{\beta}{1-\beta/(e^\beta-1)}.
\end{align}
Using this expression in Eq.~\eqref{eq:genexp-hatalpha-1} and equating the resulting expression with Eq.~\eqref{eq:genexp-hatalpha-2}, we find the function
\begin{widetext}
\begin{align}
    z(\beta) = -2n + n_\Lambda \left(1+
    \sqrt{
          1 + 
          \frac{4 \left(e^{\beta }-1\right) n}
          {\beta  n_\Lambda
            \left[
                 \meanlnx - \ln\meanx
                +\ln\left(1/\beta+1/\{1-e^{\beta }\}\right)
                \right]
         }
         }
        \right)
    \label{eq:genexp-z-of-beta}
\end{align}
\end{widetext}
the root of which gives $\hat\beta$.

To find the global maximum of the log-likelihood, one iterates over all intervals $[y_j,y_{j+1})$ checking both at the left boundary and within the interval.

\subsubsection{Special case $\beta=\alpha$}

In \cite{clauset_power-law_2009}, an additional model for non-pathological cores was introduced, namely
\begin{align}
    \label{eq:force-exp-pareto}
    p(x)=\begin{cases}
            C\exp\Big[-\alpha(x/\xm-1)\Big], & 0\leq x\leq\xm\\
            C\left(\xm/x\right)^\alpha, & x>\xm
         \end{cases}
\end{align}
(or ``forced exp-Pareto'') with normalization constant
\begin{align}
    C = \frac{\alpha(\alpha-1)}{\xm \Big((\alpha-1)e^\alpha+1\Big)}.
\end{align}
Note that for this case we have both continuity in $p(x)$ as well as its first derivative at $x=\xm$. 

The log-likelihood is given as
\begin{align}
    \ln \mathcal L = n\ln C 
                 - \alpha n_\Lambda\left(\left<\ln x\right>_\Lambda - \ln{\xm}\right)
                 - \alpha n_S\left(\frac{1}{\xm}\left<x\right>_S-1\right).
\end{align}
with $\Lambda$ and $S$ defined as above. The derivative for $\alpha$ can be found as
\begin{align}
    \frac{\partial \ln\mathcal L}{\partial \alpha} &=
    n\frac 1 \alpha + n\frac 1 {\alpha-1} - n\frac{\alpha e^\alpha}{(\alpha-1)e^\alpha+1}
    - n_\Lambda \left<\ln \frac{x}{\xm}\right>_\Lambda 
    \nonumber \\
    &\ \ - n_S \left( \frac{\left < x\right>_S}{\xm} - 1\right),
\end{align}
the zero of which can be found numerically.

For regions where the sets $\Lambda$ and $S$ are constant, the derivative by $\xm$ is given as
\begin{align}
    \frac{\partial \ln\mathcal L}{\partial \xm} = 
    -\frac{n}{\xm} +\frac{\alpha n_\Lambda}{\xm} + \frac{\alpha n_S\left< x\right>_S}{\xm^2}
\end{align}
and therefore
\begin{align}
    \hat \alpha = \frac{n \xm}{n_\Lambda + n_S\left< x\right>_S}.
\end{align}
Now we can proceed similarly to above, iterating through the ordered set of unique observation values $y_j\in Y$. For each pair of observations $(y_j, y_{j+1})$ we can find the numerical solution of the equation
\begin{align}
    \frac{\partial \ln\mathcal L}{\partial \alpha}\Big|_{\alpha=\hat\alpha} = 0
\end{align}
using the bisection method on the interval $\xm\in[y_j,y_{j+1}]$ to find $\hatxm$.

\subsection{Algebraic core}
\label{sec:alg-pareto}
\subsubsection{General case}
We define the ``alg-Pareto'' pdf as
\begin{align}
    \label{eq:alg-pareto}
    p(x)=\begin{cases}
            C\left[2-\left(x /\xm\right)^\beta\right], & 0\leq x\leq\xm\\
            C\left(\xm / x\right)^\alpha, & x>\xm
         \end{cases}
\end{align}
for $\beta>0$ and with normalization constant
\begin{align}
    C = \frac{1}{\xm}\frac{(\alpha-1)(\beta+1)}{2\alpha\beta-\beta+\alpha}.
\end{align}
Note that the pdf has a singularity at $x=0$ for $-1<\beta<0$ but is normalizable nonetheless. We discuss the special case $\beta=0$ in Sec.~\ref{sec:unipareto}.

The log-likelihood is given as 
\begin{align}
    \label{eq:powpareto-logL}
    \ln\mathcal L &= n\ln C -\alpha n_\Lambda
     \left<\ln\left(\frac{x}{\xm}\right)\right>_\Lambda
     +n_S
    \left<\ln\left[2-\left(\frac{x}{\xm}\right)^\beta\right]\right>_S,
\end{align}
In order to find the value of $\alpha$ that maximizes $\mathcal L$ given $\xm$ and $\beta$, we compute
\begin{align}
    \frac{\partial}{\partial \alpha}\ln\mathcal L
        &= -n_\Lambda \left<\ln\left(\frac{x}{\xm}\right)\right>_\Lambda  
           + n \frac{\beta+1}{(\alpha-1)(2\alpha\beta-\beta+\alpha)}.
\end{align}
The zero of this expression gives us our estimator for the exponent
\begin{align}
    \label{eq:hatalpha}
    \hat \alpha &=  \frac{3\beta+1}{4\beta+2}
            +\frac{1}{4\beta+2}\sqrt{\frac{(1+\beta)(4+\lambda+\beta(8+\lambda))}{\lambda}},\ \mathrm{with}\\
    \lambda&= \frac {n_\Lambda}{n}\left<\ln\left(\frac x \xm\right)\right>_\Lambda
\end{align}
Note that $(3\beta+1)/(4\beta+2)<1$ for $\beta>-1$, which is why we disregard the second solution of the above equation that would give us exponents $\hat \alpha<1$, i.e. in a parameter regime we exclude.

Next, we want to find the maximum-likelihood estimator for $\xm$ for support regions where $\Lambda$ and $S$ are constant. Then, we find
\begin{align}
    \frac{\partial}{\partial \xm}\ln\mathcal L
        &= -\frac{n}{\xm}+\frac{\alpha n_\Lambda}{\xm}   
           + \frac{n_S}{\xm}
    \left<\frac{1}{2(\xm/x)^\beta-1}\right>_S,
\end{align}
and thus $\hatxm$ is determined by the equation
\begin{widetext}
\begin{align}
    \left<\frac{1}{2(\hatxm/x)^\beta-1}\right>_S
-
    \frac{n}{n_S}-\frac{n_\Lambda}{n_S} 
    \left(\frac{3\beta+1}{4\beta+2}
            +\frac{1}{4\beta+2}\sqrt{\frac{(1+\beta)(4+\lambda+\beta(8+\lambda))}{\lambda}}\right)
    = 0
    \label{eq:xmin_estimator}
\end{align}
\end{widetext}
where $\hat\alpha$ is given by Eq.~\eqref{eq:hatalpha}. Now we can iterate over intervals $\xm\in[y_j,y_{j+1})$ to find the maximum either at interval boundaries or within, using the bisection method.

Unfortunately, for varying $\beta$ we have to resort to maximizing Eq.~\eqref{eq:powpareto-logL} numerically. Since we do have a method to find $\hatxm$ and $\hat\alpha$ given $\beta$, we can use a one-dimensional Nelder--Mead method to find $\hat\beta$.

\subsubsection{Special case $\beta=\alpha$}

The ``forced alg-Pareto'' pdf is given as
\begin{align}
    \label{eq:force-alg-pareto}
    p(x)&=\begin{cases}
            C\left[2-\left( x /\xm\right)^\alpha\right], & 0\leq x\leq\xm\\
            C\left(\xm/x\right)^\alpha, & x>\xm
         \end{cases}\\
    C &= \frac{1}{\xm}\frac{\alpha^2-1}{2\alpha^2}.
\end{align}
Note that for this case we have both continuity in $p(x)$ as well as its first derivative at $x=\xm$. 

The log-likelihood is given by
\begin{align}
    \ln\mathcal L &= -n\ln\xm + n\ln(\alpha^2-1) - n (\ln 2 + 2\ln\alpha)
         \nonumber \\
    &\ \ 
             -\alpha n_\Lambda\left<\ln\left(\frac{x}{\xm}\right)\right>_\Lambda
     +n_S
    \left<\ln\left[2-\left(\frac{x}{\xm}\right)^\alpha\right]\right>_S.
\end{align}
The derivative by $\alpha$ (for pre-determined $\xm$) is 
\begin{align}
\frac{\partial \ln \mathcal L }{\partial \alpha } &= 
- n_{\Lambda}\left<\ln\left(\frac{x}{\xm}\right)\right>_\Lambda 
+ \frac{2n\alpha}{\alpha^2-1}
- \frac{2n}{\alpha}
\nonumber \\
&\ \ - n_{S} \left<\frac{ \ln{\left( x/{\xm} \right)}}{2\left({x}/{\xm}\right)^{-\alpha}  - 1}\right>_S.
\end{align}
The zero of this equation can be found using the Newton--Raphson method.
Since, again, the sets $\Lambda$ and $S$ are constant on the interval $\xm\in[y_j,y_{j+1})$, one may proceed by iterating through all data intervals and finding the maximum within each interval with the Nelder--Mead method.

\subsection{Uniform core}
\label{sec:unipareto}
We define the ``uni-Pareto'' pdf as
\begin{align}
    \label{eq:uni-pareto}
    p(x) &= \begin{cases}
        C & x \leq \xm\\
        C(\xm/x)^\alpha & x > \xm
    \end{cases}\\
    C &= \frac{\alpha-1}{\alpha}\frac{1}{\xm}.
\end{align}
We compute the log-likelihood as
\begin{align}
    \ln \mathcal L =   n \ln\left(\frac{\alpha-1}{\alpha}\frac{1}{\xm}\right) 
                     - \alpha n_\Lambda \left<\ln \frac x \xm\right >_\Lambda
\end{align}
with derivative
\begin{align}
    \frac{\partial}{\partial\alpha} \ln \mathcal L &=
        n\left(\frac 1 {\alpha-1} - \frac 1 \alpha\right) - n_\Lambda \left<\ln \frac{x}{\xm}\right>_\Lambda
\end{align}
for constant $\xm$. Demanding $\partial\ln\mathcal L/\partial \alpha = 0$ and $\alpha > 1$ gives us the estimator
\begin{align}
    \hat \alpha = \frac 12 + \sqrt{
                 \frac 14 + \frac{n}{n_\Lambda\left<\ln \frac{x}{\xm}\right>_\Lambda}
    }.
\end{align}


For regions where $\Lambda$ and $S$ are constant, we find
\begin{align}
    \frac{\partial}{\partial\xm} \ln \mathcal L &=
        -n \frac 1 \xm + \alpha n_\Lambda \frac 1 \xm,
\end{align}
so we have to solve
\begin{align}
 0 &= \frac{n}{(\alpha-1)\alpha} - n_\Lambda \left<\ln x\right>_\Lambda + n_\Lambda \ln\xm \\
 0 &= \frac{\alpha n_\Lambda - n}{\xm}.
\end{align}
Then
\begin{align}
  \hat \alpha = \frac{n}{n_\Lambda}
\end{align} solves the second equation. Putting this solution in the first equation gives
\begin{align}
    \ln \hatxm &= \left<\ln x\right>_\Lambda
                               -\frac{n_\Lambda}{n-n_\Lambda}.
\end{align}
The determinant of the Hessian of this problem gives
\begin{align}
    \mathrm{det}(H) = -n_\Lambda^2\exp\Big(\dots\Big),
\end{align}
which, for a two-dimensional problem like this, means that the only extremum $(\hat a, \hatxm)$ is a saddle point, i.e.~one cannot find a better maximum than given by $\xm\in Y$, which in turn means that iterating over unique observation values and then finding $\hatxm=y_j$ that maximizes $\ln\mathcal L$ suffices.

\section{Discussion and conclusion}

We discussed how to fit three classes of piece-wise Pareto-like distributions to data using maximum-likelihood estimation, with the distributions reaching non-zero and finite values for support region $0\leq x\leq \xm$. 

The results presented in this study are neither particularly insightful nor exciting. Nonetheless, they might be of use to future analyses dealing with data that is distributed according to any of the proposed shapes.

In the future and if the use case demands it, the results may be extended to discrete random variates.

Furthermore, in a future analysis, the principle of splitting observations into sets of values below and above a threshold $\xm$ might be used for efficiently minimizing the Kullback-Leibler divergence between piece-wise finite-core models and data if the data is only accessible in binned form. 

\begin{acknowledgments}
BFM expresses his gratitude to Antonio Desiderio, Sune Lehmann, and Aaron Clauset for helpful comments and discussions regarding this manuscript. BFM received funding through Grant CF20-0044, HOPE:\ How Democracies Cope with Covid-19 from the Carlsberg Foundation.
\end{acknowledgments}

%

\appendix


\section{Properties of the alg-Pareto distribution}

The mean is finite for $\alpha>2$ and given as 
\begin{align}
    \left<x\right> &= 
    \xm \frac{\left(\alpha - 1\right) \left(\beta + 1\right) \left(\beta + \left(\alpha - 2\right) \left(\beta + 1\right) + 2\right)}{\left(\alpha - 2\right) \left(\beta + 2\right) \left(2 \alpha \beta + \alpha - \beta\right)}.
\end{align}
The second moment is finite for $\alpha>3$ and given as
\begin{align}
 \left<x^2\right> &=\xm^{2}   \frac{ \left(\alpha - 1\right) \left(\beta + 1\right) \left(3 \beta + \left(\alpha - 3\right) \left(2 \beta + 3\right) + 9\right)}{3 \left(\alpha - 3\right) \left(\beta + 3\right) \left(2 \alpha \beta + \alpha - \beta\right)}.
\end{align}
The cdf is
\begin{align}
    F{\left(x \right)} = \begin{cases}
   \frac{x}{\xm} \left(2 \beta + 2 - \left(\frac{x}{\xm}\right)^{\beta }\right)\frac{\alpha - 1 }{2 \alpha \beta + \alpha - \beta} & \text{for}\:  x \leq \xm \\
    1 - \frac{  \beta + 1}{2 \alpha \beta + \alpha - \beta}\left(\frac{\xm}{x}\right)^{\alpha - 1} & \text{otherwise} \end{cases}.
\end{align}
The inverse of the cdf is
\begin{align}
F^{-1}(q) &= \xm \left(\frac{\beta +1}{(1-q) (2 \alpha  \beta +\alpha -\beta )}\right)^{1/({\alpha
   -1})}\\
   \mathrm{for\ } q &\geq \frac{(\alpha -1) (2 \beta +1)}{2 \alpha  \beta +\alpha -\beta }.
\end{align}
For values of $q$ smaller than the critical probability, the inverse has to be found as the root of the function
\begin{align}
    \varphi(x) =  -q+\frac{x}{\xm} \left(2 \beta + 2 - \left(\frac{x}{\xm}\right)^{\beta }\right)\frac{\alpha - 1 }{2 \alpha \beta + \alpha - \beta}
\end{align}
with derivative
\begin{align}
    \varphi'(x) = \left[2-\left(\frac{x}{\xm}\right)^{\beta }\right] \frac{(\alpha -1) (\beta +1) }
    {\xm (2 \alpha  \beta +\alpha -\beta )}.
\end{align}
The inverse of the cdf can be used to find median and percentiles, and for sampling from the distribution by setting random variates $X=F^{-1}(q) = q$ with $q\sim \mathcal U(0,1)$ being a uniform random variate on the interval $[0,1)$.

\section{Properties of exp-Pareto distribution}
The mean is finite for $\alpha>2$ and given as
\begin{align}
    \left<x\right> = \xm \left(\frac{\alpha  (2-\alpha +\beta)-\beta }{(\alpha -2) \left((\alpha -1) (e^{\beta}-1) +\beta \right)}+\frac{1}{\beta }\right)
\end{align}
The second moment  finite for $\alpha>3$ and given as
\begin{align}
    \left<x^2\right> = 
\frac{(\alpha -1) y^2 \left(\alpha  \left(-\beta  (\beta +2)+2 e^{\beta }-2\right)-6
   e^{\beta }+\beta  (\beta  (\beta +3)+6)+6\right)}{(\alpha -3) \beta ^2 \left((\alpha
   -1) (e^{\beta }-1)+\beta\right)}.
\end{align}
\begin{widetext}  
The cdf is
\begin{align}
F{\left(x \right)} = \begin{cases}
\frac{(\alpha -1) e^{\beta } }{(\alpha -1)
   \left(e^{\beta }-1\right)+\beta } \left(1-e^{-\beta  x/\xm}\right)& \text{for}\: x \leq \xm \\
1 - \frac \beta {\beta + (\alpha-1) (e^\beta-1)} \left(\frac\xm x\right)^{\alpha-1}& \text{otherwise}.
\end{cases}
\end{align}

The inverse of the cdf is
  
\begin{align}
F^{-1}(q) = \begin{cases}
\xm \left(\frac{(1-q) \left((\alpha -1) e^{\beta }-\alpha +\beta +1\right)}{\beta
   }\right)^{1/{(1-\alpha )}}
   & q \leq 1-\frac{\beta }{(\alpha -1) (e^{\beta }-1)+\beta}\\
-\frac{\xm}{\beta } \ln\left(1+q \left(\frac{e^{-\beta } (\alpha -\beta -1)}{\alpha
   -1}-1\right)\right)
   & \text{otherwise}
\end{cases}
\end{align}
\end{widetext}
and can be used to find median and percentiles, and for sampling from the distribution by setting random variates $X=F^{-1}(q) = q$ with $q\sim \mathcal U(0,1)$ being a uniform random variate.

\section{Properties of the pow-Pareto distribution}
The mean is finite for $\alpha>2$ and given as
\begin{align}
    \left<x\right> = \xm
\frac{2 (\alpha -1) \alpha  (\beta +1) }{\left(\alpha ^2-4\right) (\alpha +\beta )}.
\end{align}
The second moment is finite for $\alpha>3$ and given as
\begin{align}
    \left<x^2\right> = \xm^2
\frac{(\alpha -1) (\beta +1) }{(\alpha -3) (\beta +3)}
\end{align}
The cdf is
\begin{align}
F{\left(x \right)} = \begin{cases}
\frac{\alpha-1}{\alpha+\beta} \left(\frac{x}{\xm}\right)^{\beta+1}
& \text{for}\: x \leq \xm \\
1-\frac{\beta+1 }{\alpha +\beta}\left(\frac{\xm}{x}\right)^{\alpha-1}& \text{otherwise}.
\end{cases}
\end{align}

The inverse of the cdf is
\begin{align}
F^{-1}(q) = \begin{cases}
\xm \left(q\frac{\alpha+\beta}{\alpha-1}\right)^{\frac 1{\beta+1}}
& q \leq (\alpha-1)/(\alpha+\beta)\\
\xm \left(\frac{\beta+1}{(1-q)(\alpha+\beta)}\right)^{\frac 1{\alpha-1}}
   & \text{otherwise}
\end{cases}
\end{align}
and can be used to find median and percentiles, and for sampling from the distribution by setting random variates $X=F^{-1}(q) = q$ with $q\sim \mathcal U(0,1)$ being a uniform random variate.

\section{MLE estimators for a generative contact distribution model: The Santa Fe model}

The following is not incredibly related to the rest of this paper other than that it deals with fitting a model that comes with a heavy tail and a finite core, but we still think it should be written down somewhere so it will not get lost in the nirvana of tex-files, and this appendix does not feel like the worst place to do it.

A generative model to obtain long-tail contact distributions was proposed in~\cite{santafe}. As the cited work was the result of a group effort during the Santa Fe complex system's summer school 2018, we will refer to the model as the ``Santa Fe'' model hereinafter.
In this modeling framework, every individual is associated with a normally distributed propensity $Z_i\sim\mathcal N(0,1)$ to form contacts. Additionally, every possible connection $(i,j)$ is associated with a normally distributed contact propensity $Y_{ij}\sim\mathcal N(0,1)$. Then, we can construct a mixed-effect $(i,j)$-specific contact propensity as
\begin{align}
    X_{ij} = \sqrt{\rho}(Z_i + Z_j) + \sqrt{1-2\rho} Y_{ij},\qquad 0\leq \rho<1/2.
\end{align}
The first term of this equation implies that the propensity for a single contact to be established between two nodes can potentially be influenced by the individuals' respective propensities to build contacts at all. The influence of individual contact propensities can be controlled with the parameter $\rho$. In~\cite{santafe} it was shown that even though the $X_{ij}$ are normally distributed, thresholding them by a value $t$ that determines whether or not a contact is actually established can lead to a heavy-tailed contact distribution of asymptotic form
\begin{align}
    p_k &= \frac{1}{N-1} \sqrt{\frac{1-\rho}{\rho}}
    \mathrm{exp}
    \left[-\left(\frac{1-2\rho}{2\rho}\right)y_k^2
    +ty_k\left(\frac{\sqrt{1-\rho}}{\rho}\right)
    -\frac{t^2}{2\rho}
    \right],\\
    y_k &= \Phi^{-1}\left(1-\frac{k}{n-1}\right),
\end{align}
with $\Phi^{-1}(x)$ being the inverse cdf of the normal distribution $\mathcal N(0,1)$. Note that this approximation is valid for $N\gg1$ individuals in the system and contacts per individuals $k>0$ (even though the model perfectly allows for $k=0$---only the approximation fails).

While a method to fit degree distributions was presented in~\cite{santafe}, it only relies on fixing $t$ and $\rho$ such that the mean and variance of the underlying data match those of the theoretical values of the model. Since the variance is not very resilient with regard to fluctuations in the tail, we want to take a maximum-likelihood approach to obtain $t$ and $\rho$ instead.

Given a set of observations $k_i\in\Omega$, the log-likelihood to observe the data given the model is 
\begin{align}
    \ln\mathcal L &= - n \ln{\left(N - 1 \right)} -n t \yk \frac{ \sqrt{1 - \rho}}{\rho} - n\yksq \frac{ 1-2\rho}{2\rho} 
    \nonumber \\
    &\ \ + \frac{n}{2} \big( \ln{\left(1 - \rho \right)- \ln{\left(\rho \right)}}\big) - \frac{n t^{2}}{2 \rho}.
\end{align}
Demanding $\partial \ln\mathcal L/\partial t = 0$ yields
\begin{align}
    \hat t = \yk \sqrt{1-\rho}.
\end{align}
With $\partial \ln\mathcal L/\partial \rho = 0$ we find
\begin{align}
    \hat \rho = \frac{\yksq- \yk^{2} }{\yksq- \yk^{2}  + 1}.
\end{align}
In practice, the system size $N$ depends on the experimental conditions under which the data was collected. Because the likelihood of observing large values of contact numbers (i.e. on the order of the system size) is vanishingly small for large social systems, the exact number of $N$ will not necessarily matter too much.

\end{document}